\documentclass{jfm}
\usepackage{graphicx}
\usepackage{dcolumn}
\usepackage{bm}
\usepackage{color}
\usepackage{placeins}
\usepackage{array}
\usepackage{subfigure}
\usepackage{natbib}
\usepackage{amssymb}
\usepackage{amsbsy}
\usepackage{epstopdf}

\title[$Pr$ and $Ra$ number dependence of heat transport in high $Ra$ number RB]
{Prandtl and Rayleigh number dependence of heat transport in high Rayleigh number thermal convection}

\author[Richard J.A.M. Stevens, Detlef Lohse, and Roberto Verzicco]
{Richard J.A.M. Stevens$^1$, Detlef Lohse$^1$, and  Roberto Verzicco$^{1,2}$}

\affiliation{
$^1$Department of Science and Technology and J.M. Burgers Center for Fluid Dynamics, University of Twente, P.O Box 217, 7500 AE Enschede, The Netherlands,\\
$^2$Dept. of Mech. Eng., Universita' di Roma "Tor Vergata",Via del Politecnico 1, 00133, Roma.}

\pubyear{}
\volume{}
\pagerange{}
\date{\today}

\begin{document}

\maketitle

\begin{abstract}
Results from direct numerical simulation for three-dimensional Rayleigh-B\'enard convection in samples of aspect ratio $\Gamma=0.23$ and $\Gamma=0.5$ up to Rayleigh number $Ra=2\times10^{12}$ are  presented.
The broad range of Prandtl numbers $0.5<Pr<10$ is considered. In contrast to some experiments, we do not see any increase in $Nu/Ra^{1/3}$, neither due to $Pr$ number effects, nor due to a constant heat flux boundary condition at the bottom plate instead of  constant temperature boundary conditions. Even at these very high $Ra$, both the thermal and kinetic
boundary layer thicknesses obey Prandtl-Blasius scaling.
\end{abstract}

\section{Introduction}
In Rayleigh-B\'enard (RB) convection fluid in a box is heated from below and cooled from above (\cite{ahl09}). This system is paradigmatic for turbulent heat transfer, with many applications in atmospheric and environmental physics, astrophysics, and process technology. Its dynamics is characterized
 by the Rayleigh number $Ra = \beta g \Delta L^3/(\kappa\nu)$ and the Prandtl number $Pr = \nu /\kappa$. Here, $L$ is the height of the sample, $\beta$ is the thermal expansion coefficient, $g$ the gravitational acceleration, $\Delta$ the temperature difference between the bottom and the top of the sample, and $\nu$ and $\kappa$ the kinematic viscosity and the thermal diffusivity, respectively. Almost all experimental and numerical results on the heat transfer, indicated by the Nusselt number $Nu$, agree up to $Ra\approx 2\times 10^{11}$ (see 
the review of \cite{ahl09} for detailed references) 
and are in agreement with the description of
 the Grossmann-Lohse (GL) theory (\cite{gro00,gro01,gro02,gro04}). However, for higher $Ra$ the situation is less clear. 

\begin{figure}
\centering
\subfigure{\includegraphics[width=0.99\textwidth]{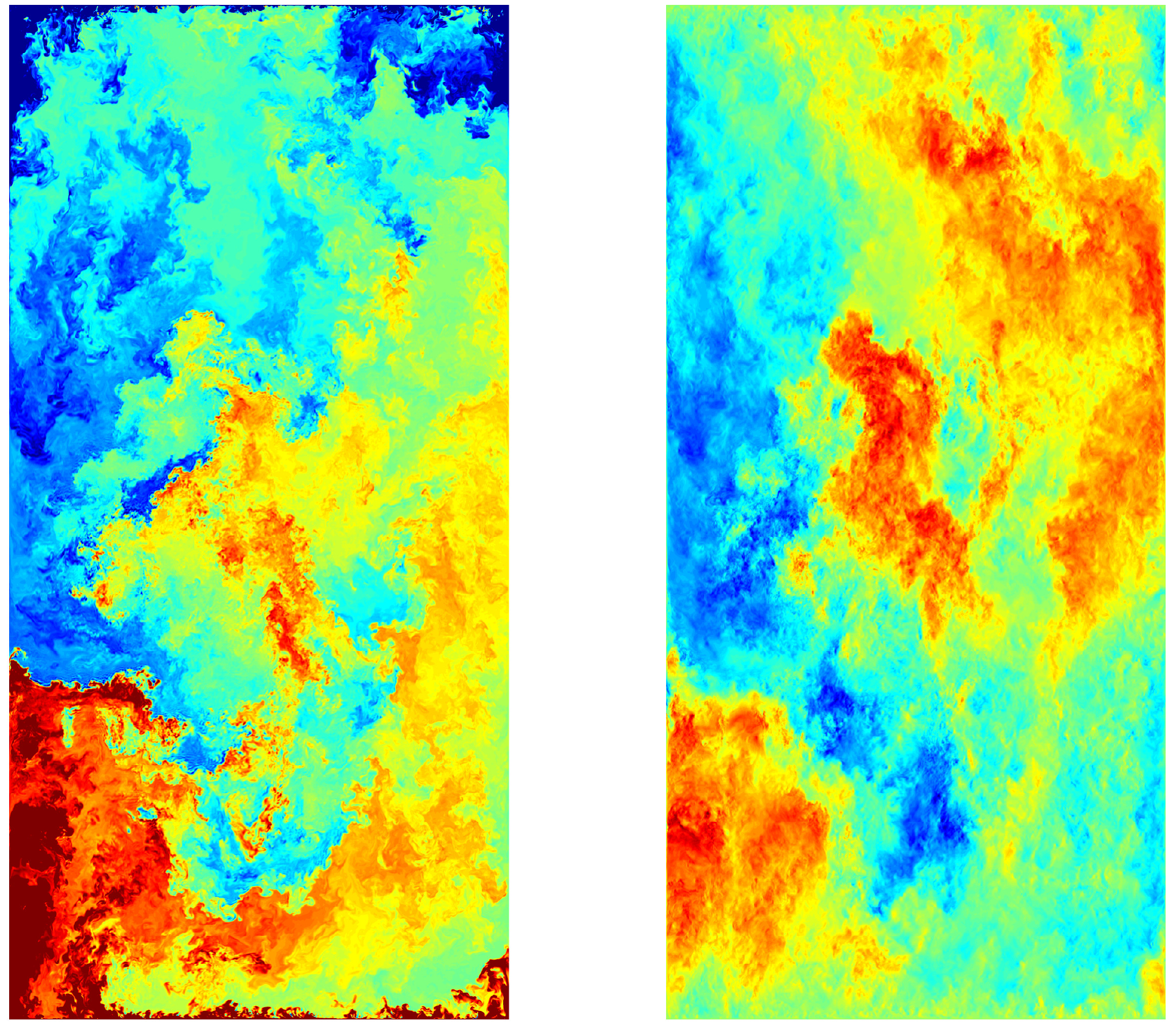}}
\caption{Visualization of the instantaneous temperature and temperature (left) and vertical velocity field (right) for the simulation at $Ra=2\times10^{12}$ and $Pr=0.7$ for $\Gamma=0.5$. Red and blue indicate warm (up flowing) and cold (down flowing) fluid, respectively, in the left (right) panel. Corresponding movies can be found as supplementary material.
}
\label{fig:figure1}
\end{figure} 

Most experiments for $Ra\gtrsim 2\times 10^{11}$ are performed in samples with aspect ratios 
$\Gamma \equiv D/L=0.5$ and $\Gamma=0.23$, where $D$ and $L$ are the diameter and height of the sample, respectively. The majority of these experiments are performed with liquid helium near its critical point 
(\cite{cha01,nie00,nie01,nie06,roc01,roc02,roc10}). 
{While \cite{nie00,nie01} and \cite{nie06} found a $Nu$ increase with $Nu \varpropto Ra^{0.31}$, the experiments by \cite{cha01} and \cite{roc01,roc02,roc10} gave a steep $Nu$ increase with $Nu \varpropto Ra^{0.38}$. In these helium experiments the $Pr$ number increases with increasing $Ra$.
 \cite{fun09} and \cite{ahl09c,ahl09d} performed measurements around room temperature with high pressurized gases with nearly constant $Pr$ and do not find such a 
 steep increase. 
 \cite{nie10b} found two $Nu \varpropto Ra^{1/3}$ branches in a $\Gamma=1$ sample. The high $Ra$ number branch is $20\%$ higher than the low $Ra$ number branch. By necessity, $Nu$ increases more steeply in the transition region. The scaling in the transition region happens to be around
 $Nu \varpropto Ra^{0.5}$.
 There are thus considerable differences in the heat transfer obtained in these different experiments in the high $Ra$ number regime. 
 Very recently, Ahlers and coworkers (see \cite{ahl10} and addendum to  \cite{ahl09d}) even found two different branches in {\it one} experiments with the steepest branch going as $Nu \varpropto Ra^{0.36}$.

\begin{figure}
\centering
\subfigure{\includegraphics[width=0.49\textwidth]{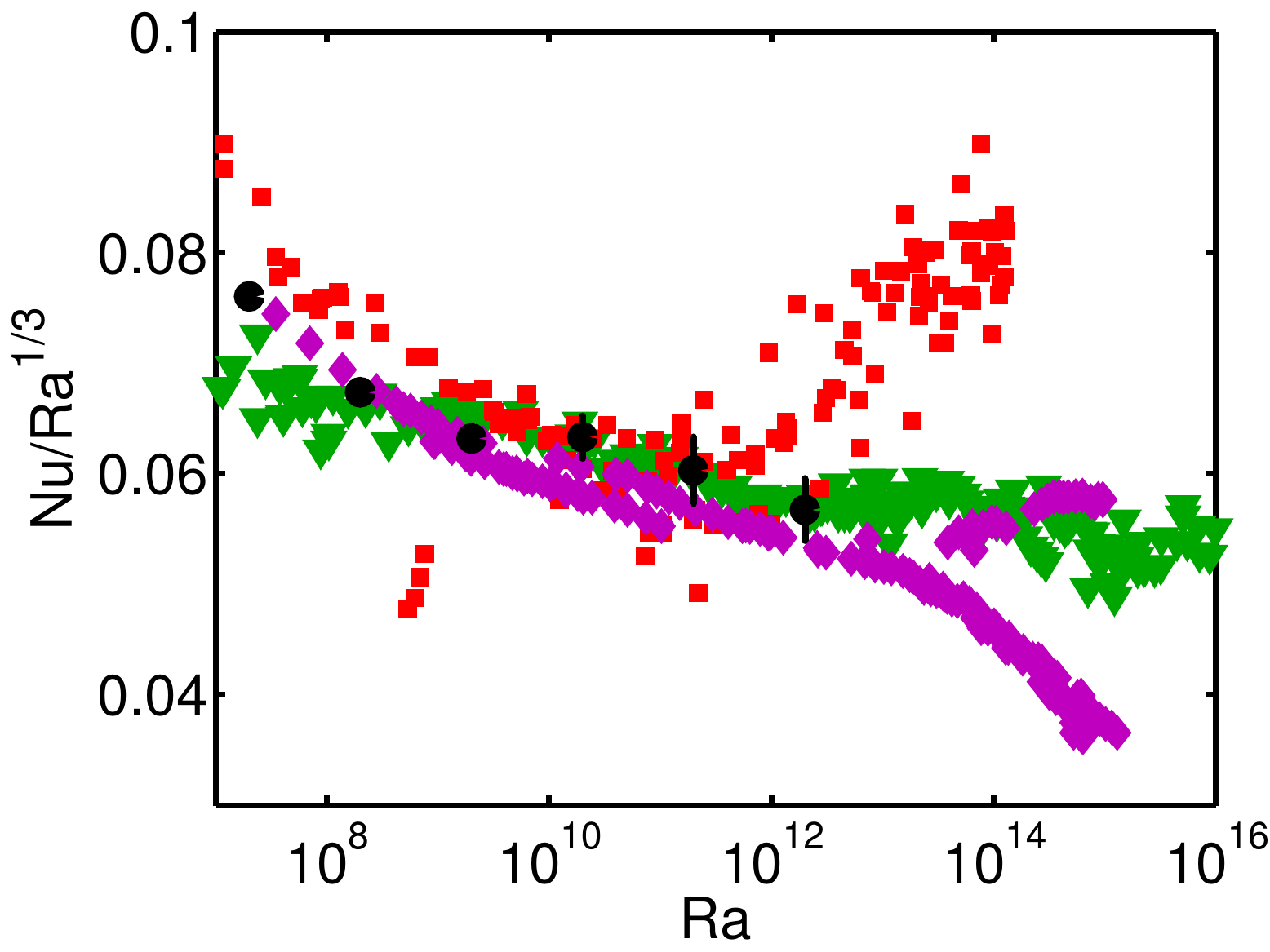}}
\subfigure{\includegraphics[width=0.49\textwidth]{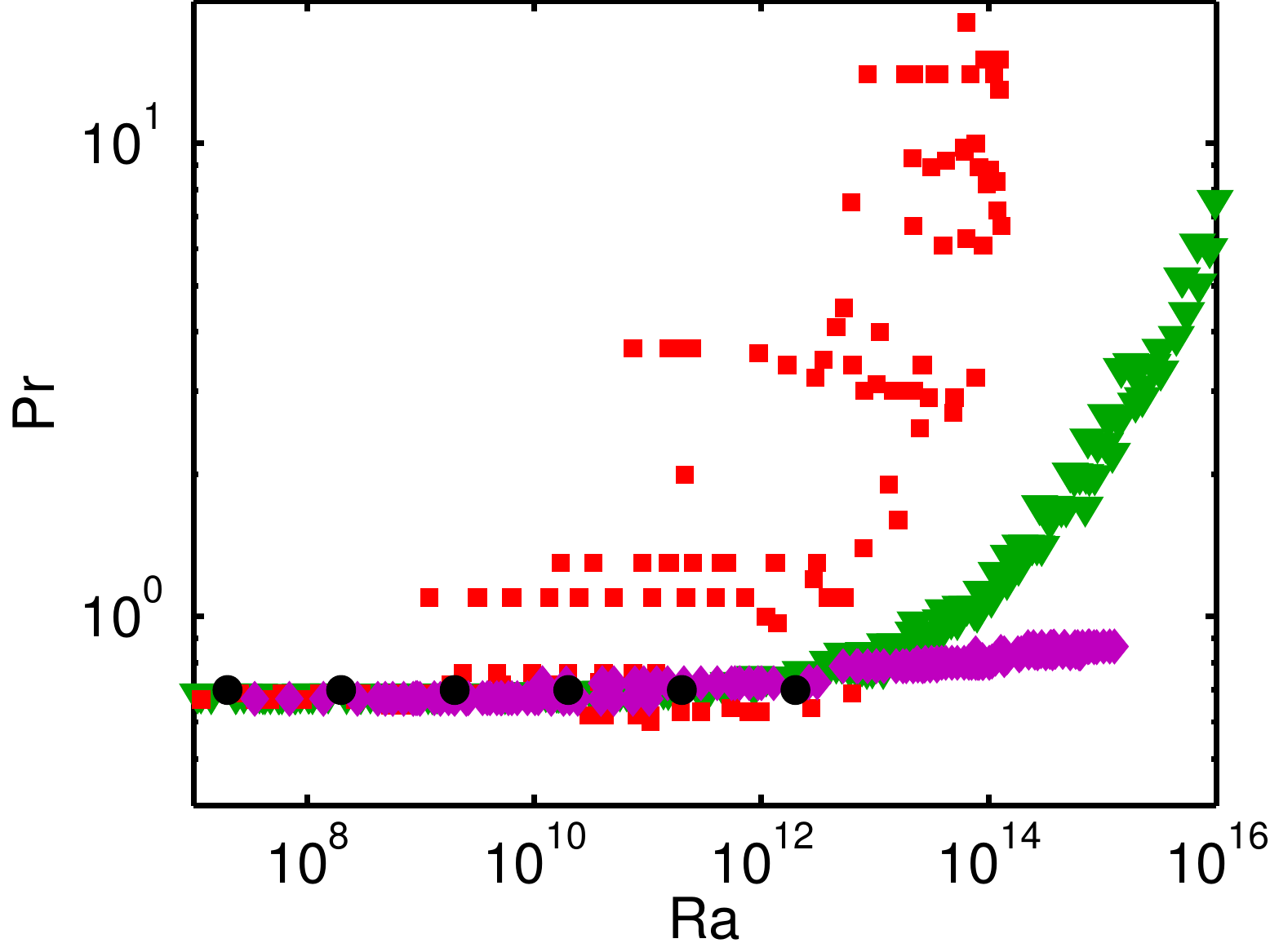}}
\caption{(a) $Nu$  vs $Ra$  for $\Gamma=0.5$. The green downward pointing triangles are experimental data from \cite{nie00,nie01} after a reanalysis reported in \cite{nie06}, the red squares are from \cite{cha01}, the purple diamonds are from \cite{fun09} and \cite{ahl09c,ahl09d,ahl10}. The DNS results for $Pr=0.7$ are indicated in black and are from \cite{ste10}. The data point for $Ra=2\times10^{12}$ is from this study. When the vertical error bar is not visible the error is smaller than the dot size. (b) Parameter space for the data presented in panel a. Symbols as in panel a.
}
\label{fig:figure2}
\end{figure}

\begin{figure}
\centering
\subfigure{\includegraphics[width=0.49\textwidth]{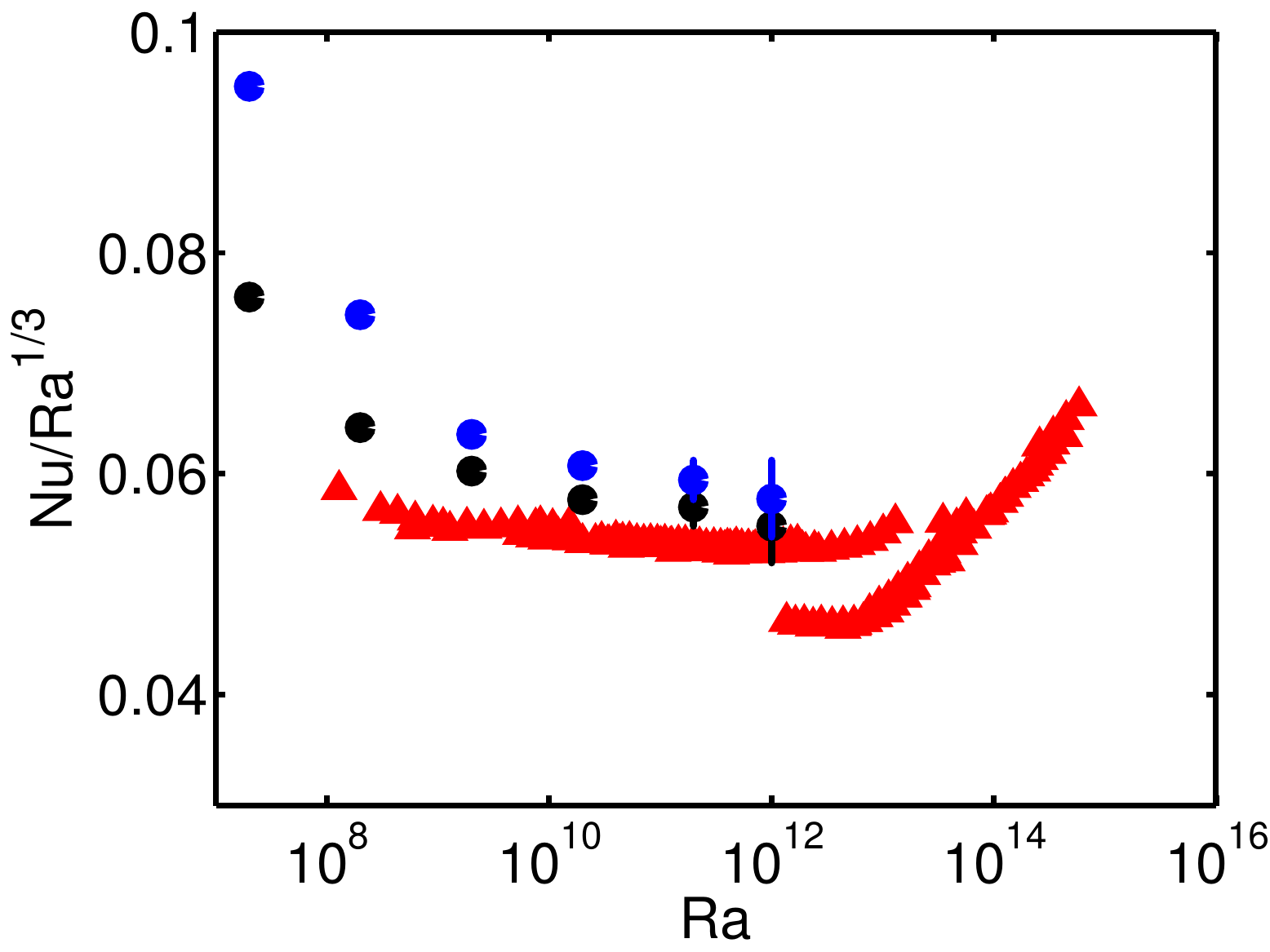}}
\subfigure{\includegraphics[width=0.49\textwidth]{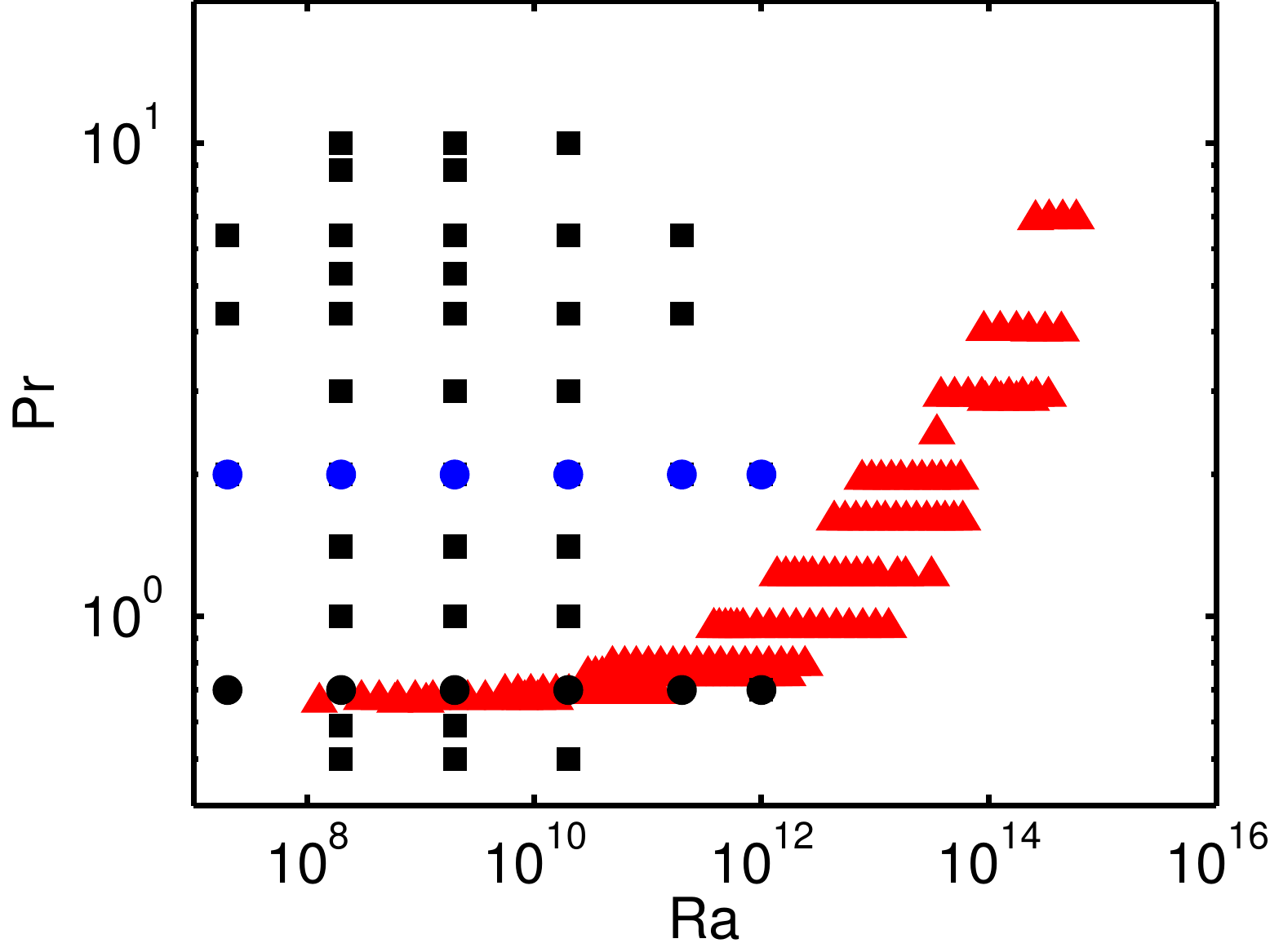}}
\caption{(a) $Nu$  vs $Ra$  for $\Gamma=0.23$. The upward pointing red triangles are from \cite{roc10}. The DNS results for $Pr=0.7$ and $Pr=2.0$ are indicated in black and blue, respectively. When the vertical error bar is not visible the error is smaller than the dot size. (b) Parameter space for the data presented in panel a. Symbols as in panel a. The black squares indicate DNS simulations presented in this paper.
}
\label{fig:figure2b}
\end{figure}

There is no clear explanation for this disagreement although it has been
 conjectured that variations of the $Pr$ number, the use of constant temperature or a constant heat flux condition at the bottom plate, the finite conductivity of the horizontal plates and side wall, non Oberbeck-Boussinesq effects, i.e. the dependence of the fluid properties on the temperature, the existence of multiple turbulent states (\cite{gro11}), and even wall roughness and temperature conditions outside the sample might play a role. Since the above differences among experiments might be induced by unavoidable technicalities in the laboratory set--ups, within this context, direct numerical simulations are the only possibility to obtain neat reference data that strictly adhere to the intended theoretical problem and that could be used as guidelines to interpret the experiments: This is the main motivation for the present study.

\section{Numerical procedure}
We start this paper with a description of the numerical procedure that is used to investigate the influence on the heat transfer for two of the issues mentioned above. First, we discuss  the effect of the $Pr$ number on the heat transport in the high $Ra$ number regime. Subsequently, we will discuss the difference between simulations performed with a constant temperature at the bottom plate with simulations with a constant heat flux at the bottom plate. We take the constant heat flux condition only at the bottom plate, because in real setups the bottom plate is in contact with a heater while the top plate is connected to a thermostatic bath. Thus, the condition of constant heat flux applies  at most only to the bottom plate 
(\cite{nie00,nie01} and \cite{nie06}).
At the top plate 
constant temperature boundary conditions are assumed to strictly hold, i.e. perfect heat transfer
to the recirculating cooling liquid.  
We will conclude the paper with a brief summary, discussion, and outlook to future simulations. 

The flow is solved by numerically integrating the  three-dimensional Navier-Stokes equations within the Boussinesq approximation. The numerical method is already described in \cite{ver97,ver03} and \cite{ver08} and here it is sufficient to mention that the equations in cylindrical coordinates are solved by central second-order accurate finite-difference approximations in space and time. We performed simulations with constant temperature conditions at the bottom plate for $2\times 10^7<Ra<1\times 10^{12}$ and $0.5<Pr<10$ in an aspect ratio $\Gamma=0.23$ sample. We also present 
 results for a simulation at $Ra=2\times10^{12}$ at $Pr=0.7$ in a $\Gamma=0.5$ sample. In addition, we performed simulation with a constant heat flux at the bottom plate and a constant temperature at the top plate, see \cite{ver08} for details, for $Pr=0.7$, $\Gamma=0.5$, and $2\times 10^6 \leq Ra \leq 2\times10^{11}$. Because in all simulations the temperature boundary conditions are precisely assigned, the surfaces are infinitely smooth,
 and the Boussinesq approximation is unconditionally valid,
 the simulations provide a clear reference case for present and future experiments.

In \cite{ste10} we investigated the resolution criteria that should be satisfied in a fully resolved DNS simulation and \cite{shi10} determined the minimal number of nodes that should be placed inside the boundary layers. The resolutions used here are based on this experience and we stress that in this study we used even better spatial resolution than we used in \cite{ste10} to be sure that the flow is fully resolved. 

To give the reader some idea of the scale} of this study we mention the resolutions that were used in the most demanding simulations, i.e., simulations that take at least 100.000 DEISA CPU hours each. For $\Gamma=0.23$ this are the simulations at $Ra=2\times10^{11}$, which are performed on either a $641 \times 185 \times 1281$ (azimuthal, radial, and axial number of nodes) grid for $Pr=0.7$ and $Pr=2.0$ or on a $769 \times 257 \times 1537$ grid for $Pr=4.38$ and $Pr=6.4$. The simulations at $Ra=1\times10^{12}$ are performed on a $1081\times301\times 2049$ grid. The simulations for $Ra=2\times10^{11}$ were run for at least $100$ dimensionless time units 
(defined as $L/\sqrt{\beta g \Delta L}$), 
while these simulations at $Ra=1\times10^{12}$ cover about $30-40$ time units. The simulation with a constant heat flux condition at the bottom plate and constant temperature condition at the top plate in a $\Gamma=0.5$ sample with $Pr=0.7$ at $Ra=2.25 \times 10^{11}$ is performed on a $1081\times 351 \times 1301$ grid. The simulation at $Ra=2\times10^{12}$ with $Pr=0.7$ in the $\Gamma=0.5$ sample has been performed on a $2701\times 671 \times 2501$ grid, which makes it the largest fully bounded turbulent flow  simulation ever. This simulation takes about $100.000$ vectorial CPU hours on HLRS (equivalent to $\approx 9 \times 10^6$ DEISA CPU hours). To store one snapshot of the field ($T$, $u_1$,$u_3$, because $u_2$ follows from continuity) costs 160 GB in binary format. A snapshot of this flow is shown in figure \ref{fig:figure1}. Movies of this simulation are included in the supplementary material.

\section{Numerical results on $Nu(Ra,Pr,\Gamma)$}
In figure \ref{fig:figure2}a and figure \ref{fig:figure2b} the DNS results for $Pr=0.7$ in the
$\Gamma=0.23$ and $\Gamma=0.5$ samples are compared with experimental data. The result for $Ra=2\times10^{12}$ in the  $\Gamma=0.5$ sample agrees well with the experimental data of  \cite{nie00,nie01}, \cite{nie06}, \cite{fun09}, and \cite{ahl09c,ahl09d}, while there is a visible difference with the results of \cite{cha01}. A comparison of the results for $\Gamma=0.23$ with the experimental data of \cite{roc10} shows that there is a good agreement for higher $Ra$ numbers, while for lower $Ra$ we obtain slightly 
larger
$Nu$ than in those experiments. We again stress that
we performed resolution checks for this $\Gamma=0.23$ case (up to $Ra=2\times10^{10}$), and in addition  
considering the good agreement with the results for $\Gamma=0.5$, 
we exclude that our DNS results overestimate $Nu$. 

\begin{figure}
\centering
\subfigure{\includegraphics[width=0.49\textwidth]{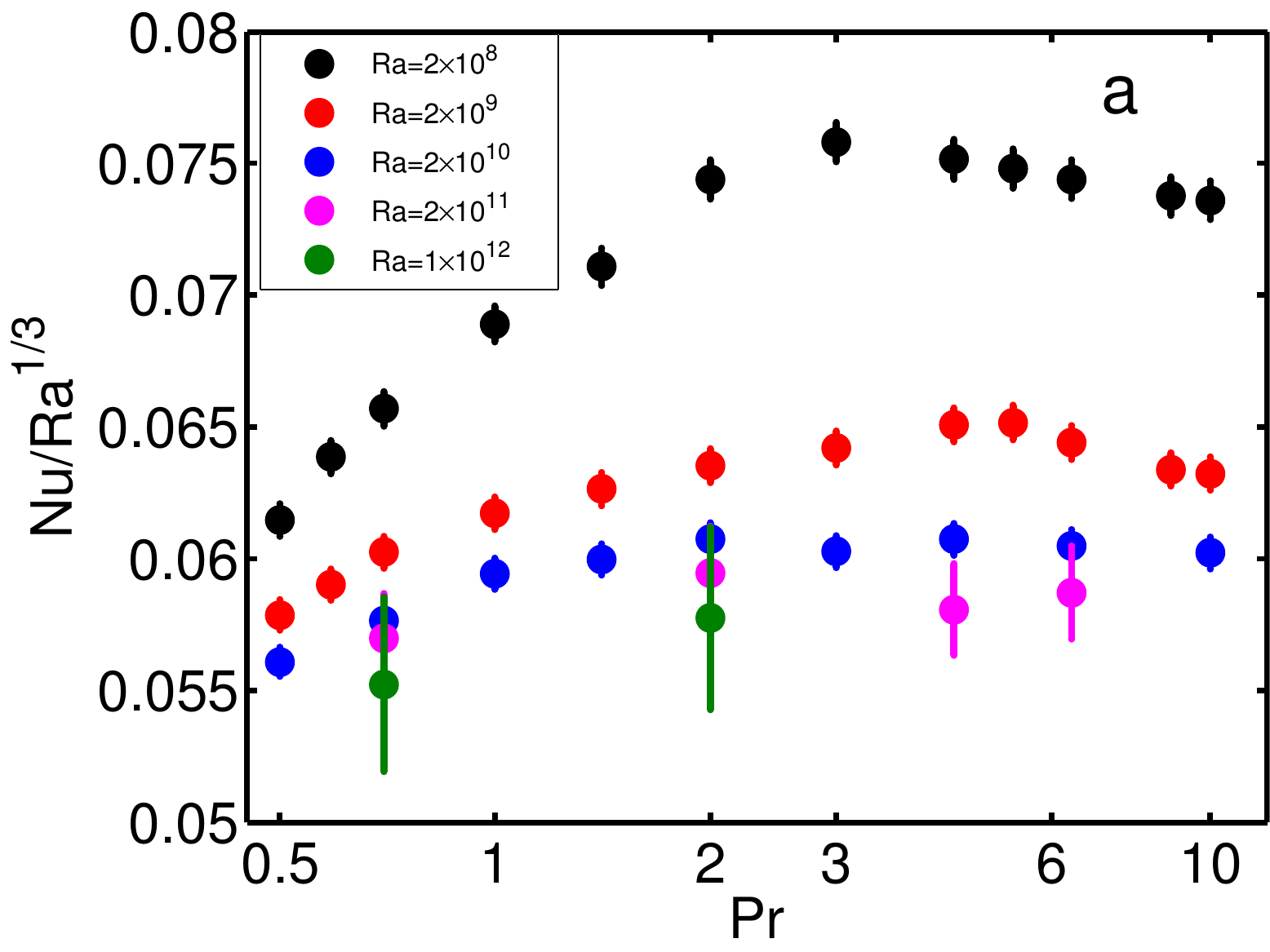}}
\subfigure{\includegraphics[width=0.49\textwidth]{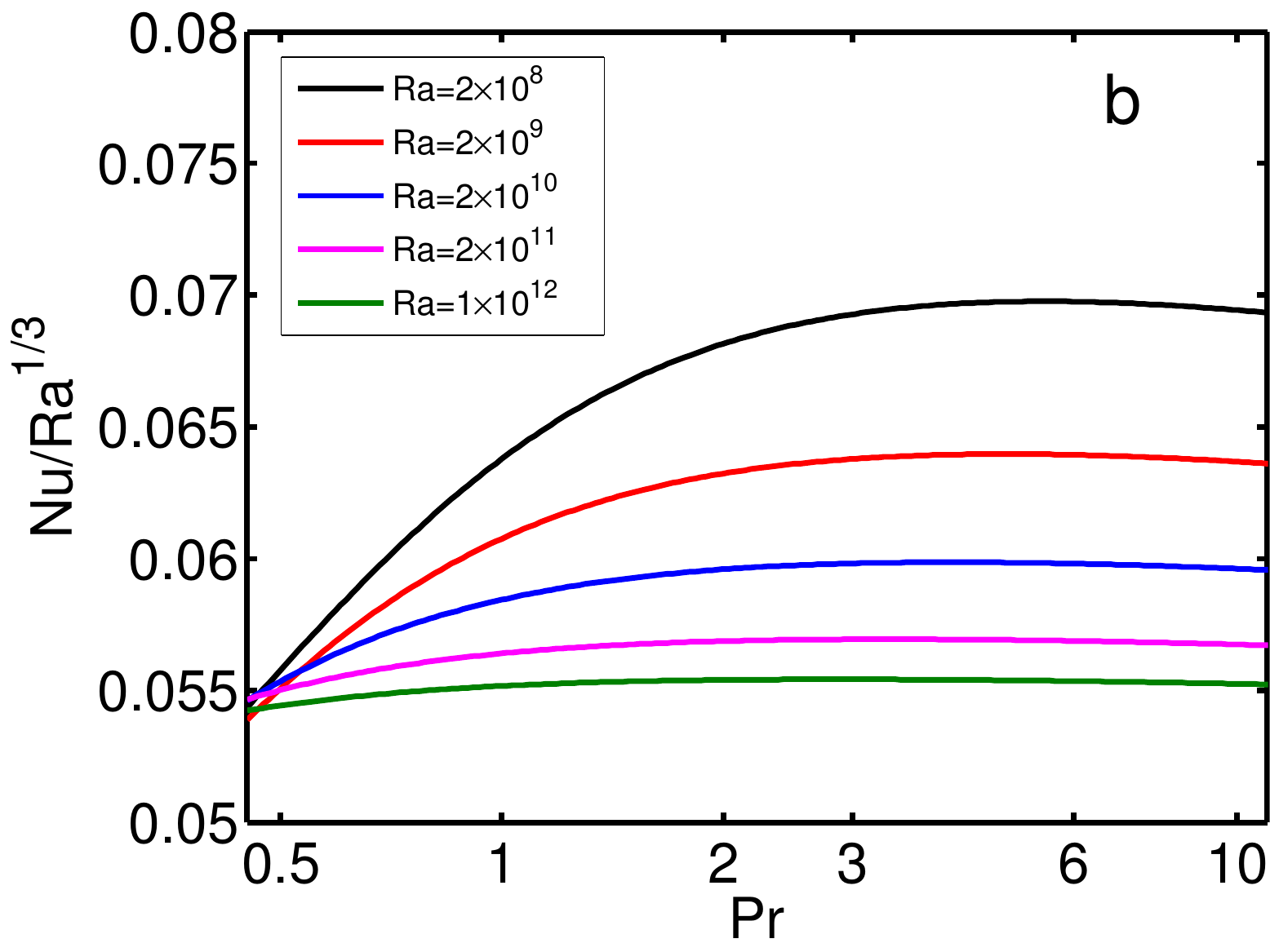}}
\caption{$Nu$  vs $Pr$. The results for $Ra=2\times10^8$, $Ra=2\times10^9$, $Ra=2\times10^{10}$, $Ra=2\times10^{11}$ and $Ra=10^{12}$ are indicated in black, red, blue, purple, and dark green, respectively. Panel a gives the DNS results for $\Gamma=0.23$ and panel b the prediction from the GL model for $\Gamma=1$.
The slight decrease of Nu for $Pr\ge 3$ and not too large Ra is also seen in experiment of \cite{ahl01} and \cite{xia02}.
}
\label{fig:figure3}
\end{figure}

Figure \ref{fig:figure2}b and \ref{fig:figure2b} show that in some experiments the $Pr$ number increases with increasing $Ra$. This difference in $Pr$ is often mentioned as one of the possible causes for the observed differences in the heat transfer between the experiments. Figure \ref{fig:figure3} shows the $Nu$ number as function of $Pr$ for different $Ra$. This figure shows that the effect of the $Pr$ number on the heat transfer decreases with increasing $Ra$. This means that the differences in the heat transport that are observed between the experiments for $Ra\gtrsim 10^{11}$, see figure \ref{fig:figure2}a and \ref{fig:figure2b}a, are not a $Pr$ number effect. This is in agreement with the theoretical prediction of the GL-model for $\Gamma=1$, which is shown in figure \ref{fig:figure3}.

\section{Scaling of thermal and kinetic boundary layers}
\begin{figure}
\centering
\subfigure{\includegraphics[width=0.49\textwidth]{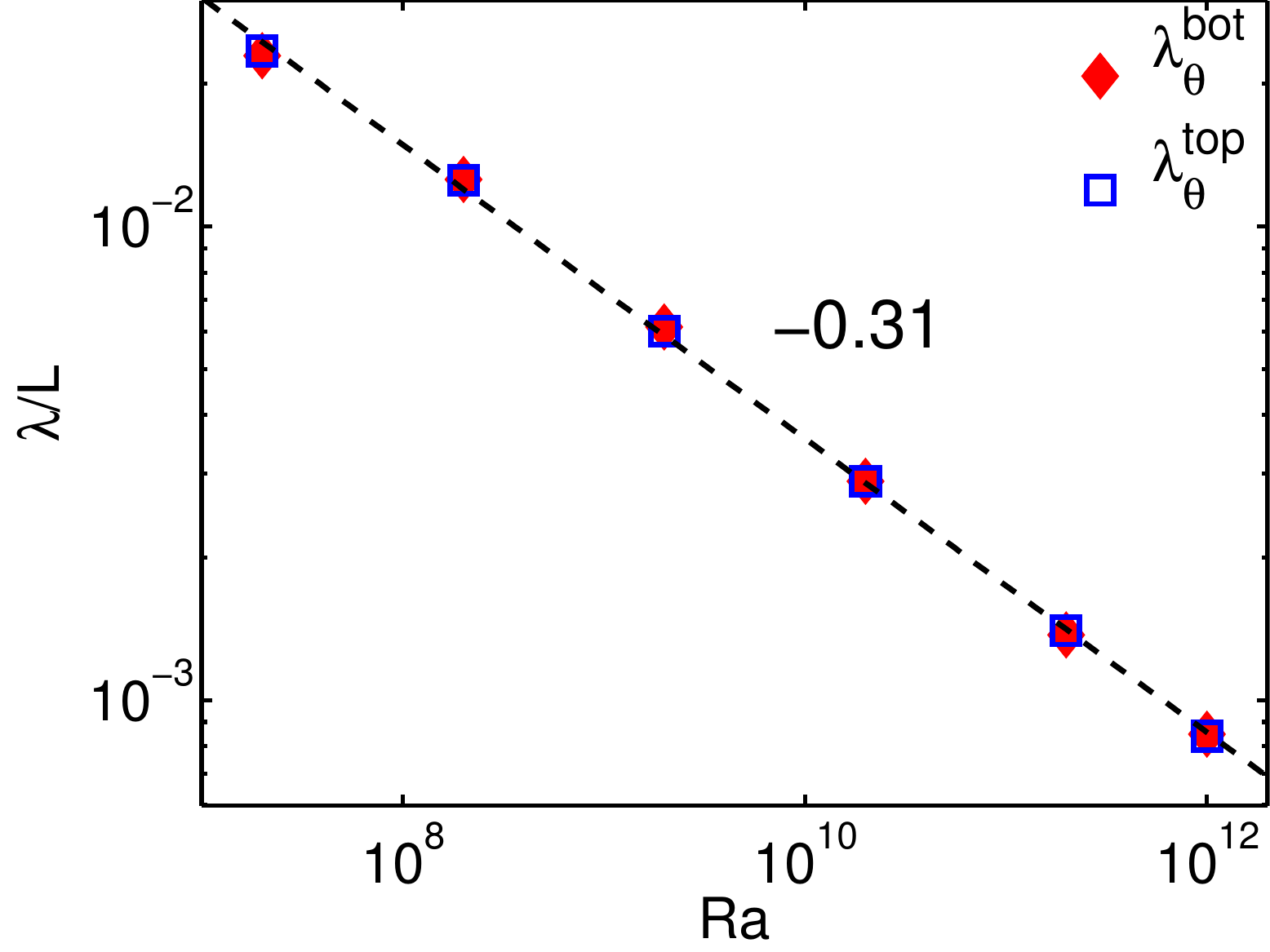}}
\subfigure{\includegraphics[width=0.49\textwidth]{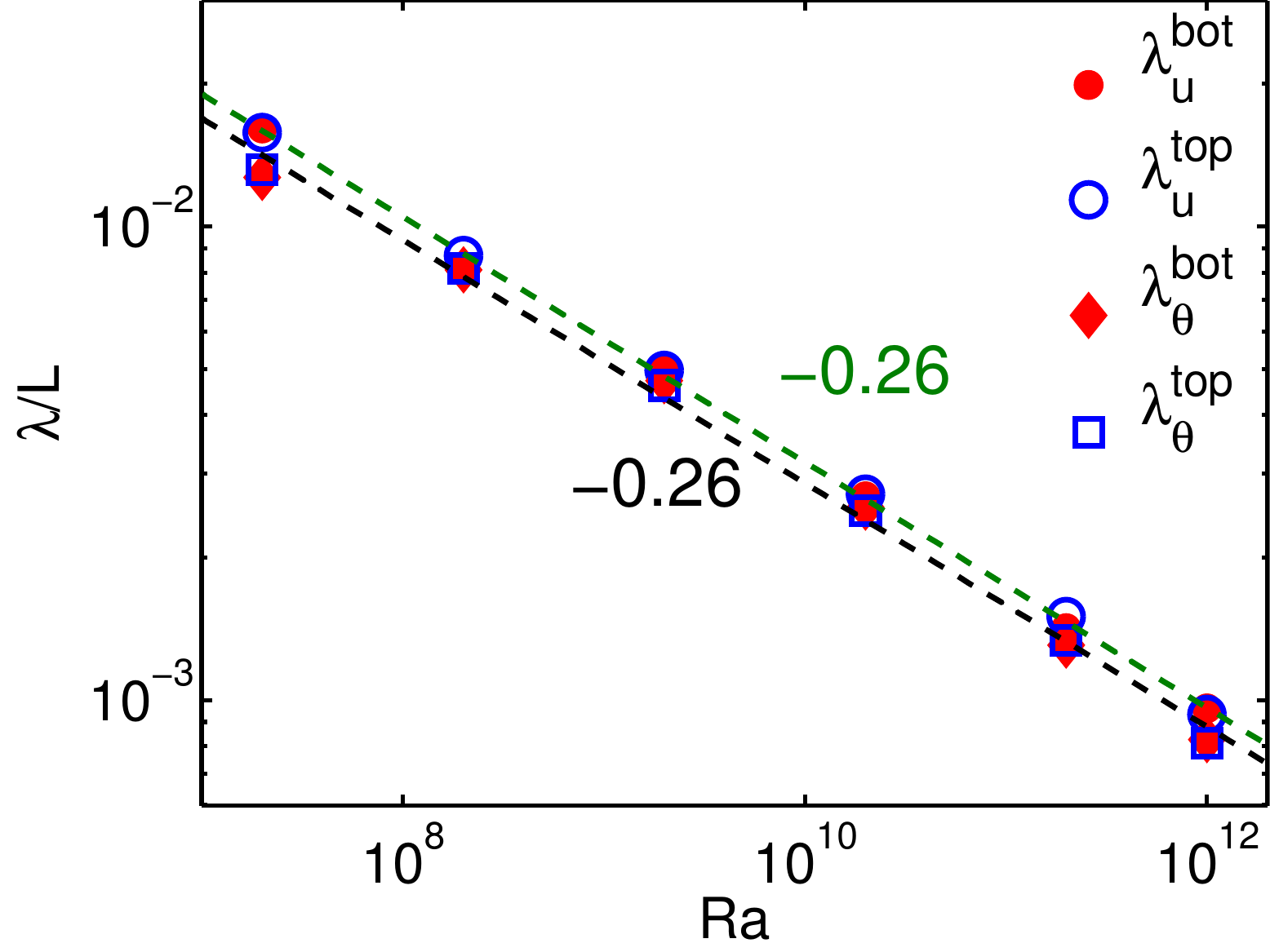}}
\caption{BL thicknesses for $Pr=0.7$ in the $\Gamma=0.23$ sample. a) The thermal BL thickness close to the bottom and top plate averaged over the entire horizontal area scales with $Ra^{-0.31}$ (black dashed line) b) The kinetic (dark green line) and thermal (black line) BL thicknesses averaged between $0.25D/2<r<0.50D/2$ scale with $Ra^{-0.26}$. Note that this is slightly less than in figure \ref{fig:figure6}, when we average $\lambda_\theta$ over the full area.
}
\label{fig:figure6}
\end{figure}

We determined the thermal and kinetic BL thickness for the simulations in the $\Gamma=0.23$ sample. The horizontally averaged thermal BL thickness ($\lambda_\theta$) is determined from $\lambda_\theta^{sl}(r)$, where $\lambda_\theta^{sl}(r)$ is the intersection point between the linear extrapolation of the temperature gradient at the plate with the behavior found in the bulk (\cite{ste10b}). In figure \ref{fig:figure6}a it is shown that the scaling of the thermal BL thickness is consistent with the $Nu$ number measurements when the horizontal average is taken over the entire plate.

The horizontally averaged kinetic BL thickness ($\lambda_u$) is determined from $\lambda_u^{"}(r)$, where $\lambda_u^{"}(r)$ is based on the position where the quantity $\epsilon_u^":= \bf{u} \cdot \nabla^2 \bf{u}$ reaches its maximum. We use this quantity as \cite{ste10} and \cite{shi10} showed that it represents the kinetic BL thickness better than other available methods. \cite{ste10b,ste10} also explained that this quantity cannot be used close to the sidewall as here it misrepresents the kinetic BL thickness. For numerical reasons, it can neither be calculated accurately in the center region.  To be on the safe side we horizontally average the kinetic BL between $0.25D/2<r<0.75D/2$, where $D$ indicates the diameter of the cell and all given values refer to that used.

\begin{figure}
\centering
\subfigure{\includegraphics[width=0.49\textwidth]{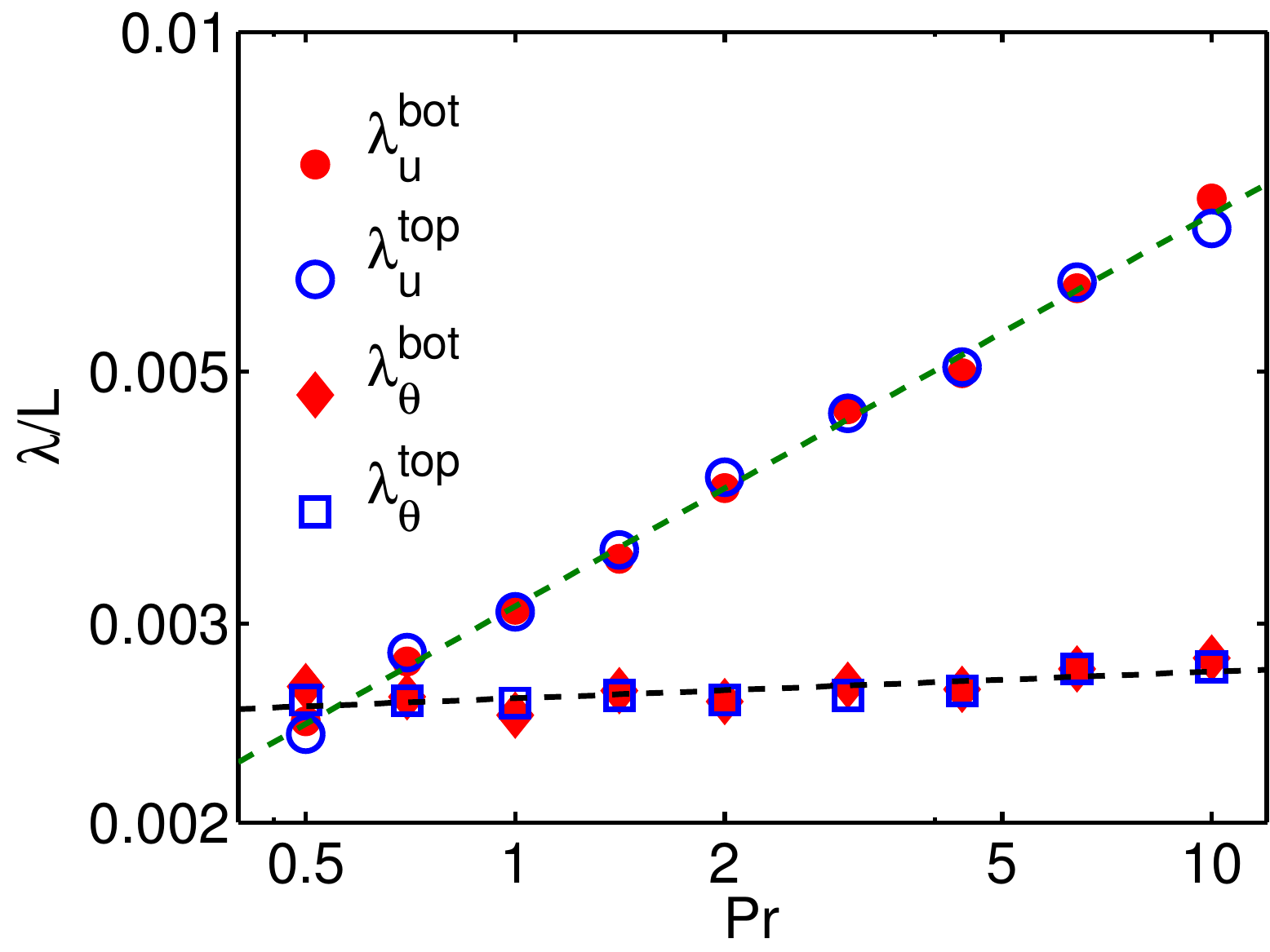}}
\subfigure{\includegraphics[width=0.49\textwidth]{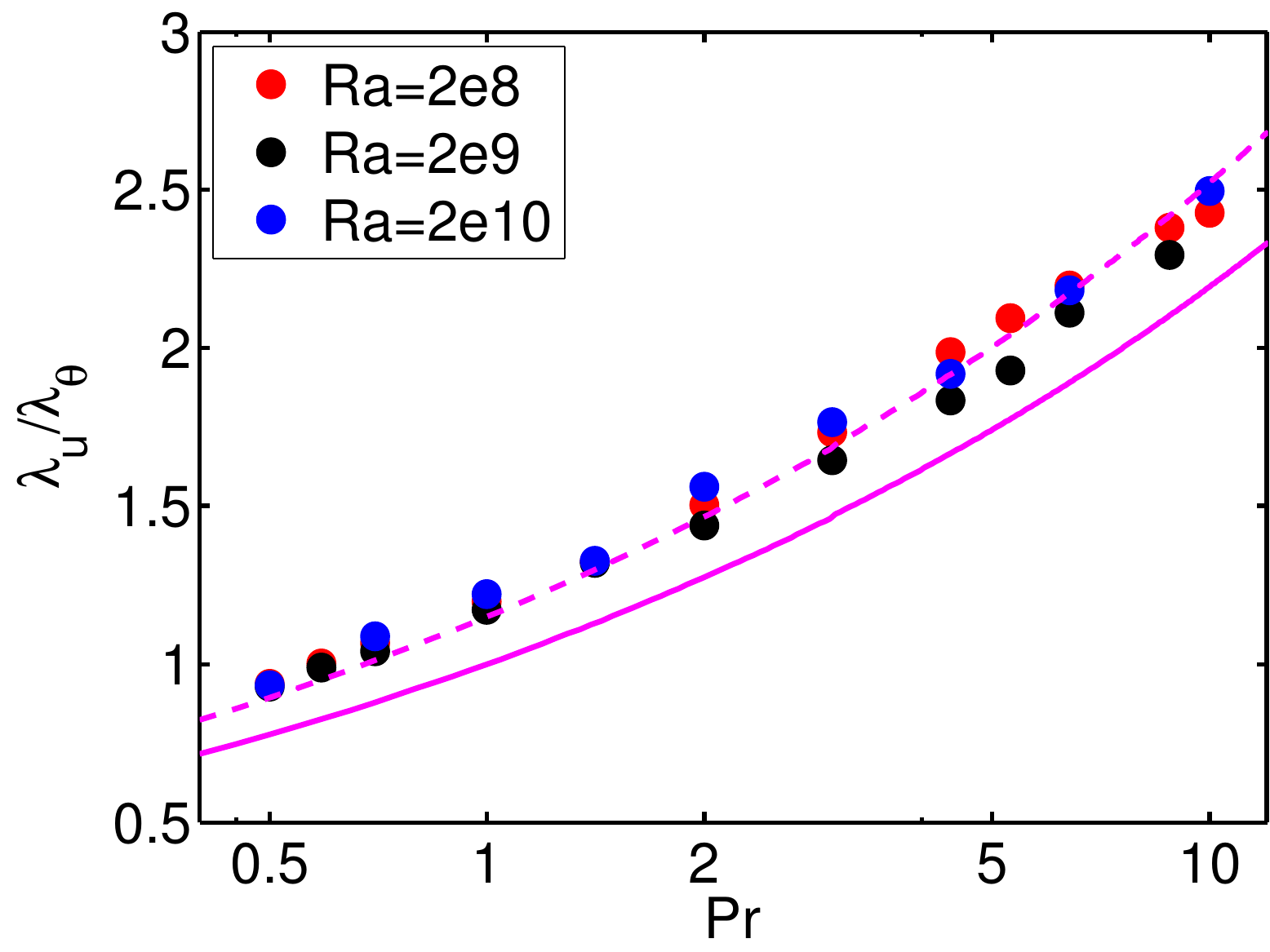}}
\caption{a) Kinetic and thermal BL thicknesses close to the bottom and top plate for $Ra=2\times10^{10}$. b) The ratio between the kinetic and thermal BL thickness as function of $Pr$ for different $Ra$. The green solid line indicates the prediction from the Prandtl-Blasius theory (\cite{shi10}). The green dashed line, which lies $15\%$ above the theoretical prediction, is a guide to the eye. All numerical data are horizontally averaged for $0.25D/2<r<0.50D/2$.}
\label{fig:figure7}
\end{figure}

Figure \ref{fig:figure6}b reveals that for $Pr=0.7$ the kinetic and thermal BL thickness have the same scaling and thickness over a wide range of $Ra$. In figure \ref{fig:figure7} the ratio $\lambda_u / \lambda_\theta$ is compared with results of the Prandtl-Blasius BL theory. We find a {\it constant} difference of about $15\%$ between the numerical results and the theoretical PB type prediction, see e.g.\ \cite{shi10}. We emphasize that the deviation of the prefactor of only 15\% is remarkably small, given that  the PB boundary layer theory has been developed  for parallel flow over an infinite flat plate, whereas  here in the  aspect ratio $\Gamma=0.23$ cell one can hardly find such  regions of parallel flows at the top and bottom plates. Nonetheless, the scaling and even the ratio of the kinetic and thermal boundary layer thicknesses  for these large $Ra$ numbers   is well described by Prandtl-Blasius BL theory. This result agrees with the experimental results of \cite{qiu98} and \cite{sun08}. Indeed, \cite{qiu98} showed that the kinetic BL near the sidewall obeys the scaling law of the Prandtl-Blasius laminar BL and \cite{sun08} showed the same for the boundary layers near the bottom plate. Recently, \cite{zho10,zho10b} have developed a method of expressing velocity profiles in the time-dependent BL frame and found that not only the scaling obeys the PB expectation, but even the rescaled velocity and temperature profiles From all this we can exclude that at $Ra\lesssim 10^{12}$ the BL is turbulent.

\section{Constant temperature versus constant heat flux condition at the bottom plate}

\begin{figure}
\centering
\subfigure{\includegraphics[width=0.49\textwidth]{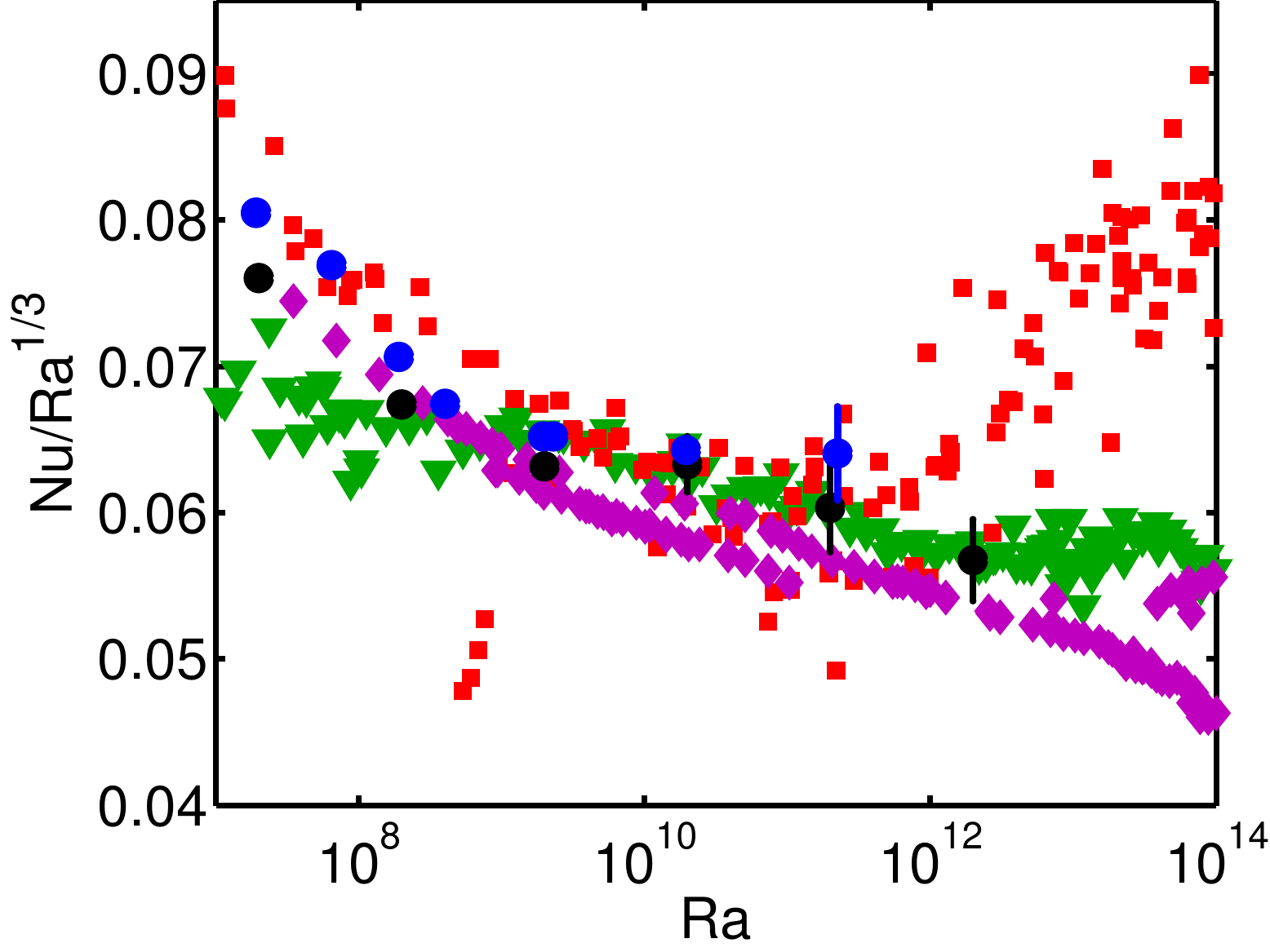}}
\subfigure{\includegraphics[width=0.49\textwidth]{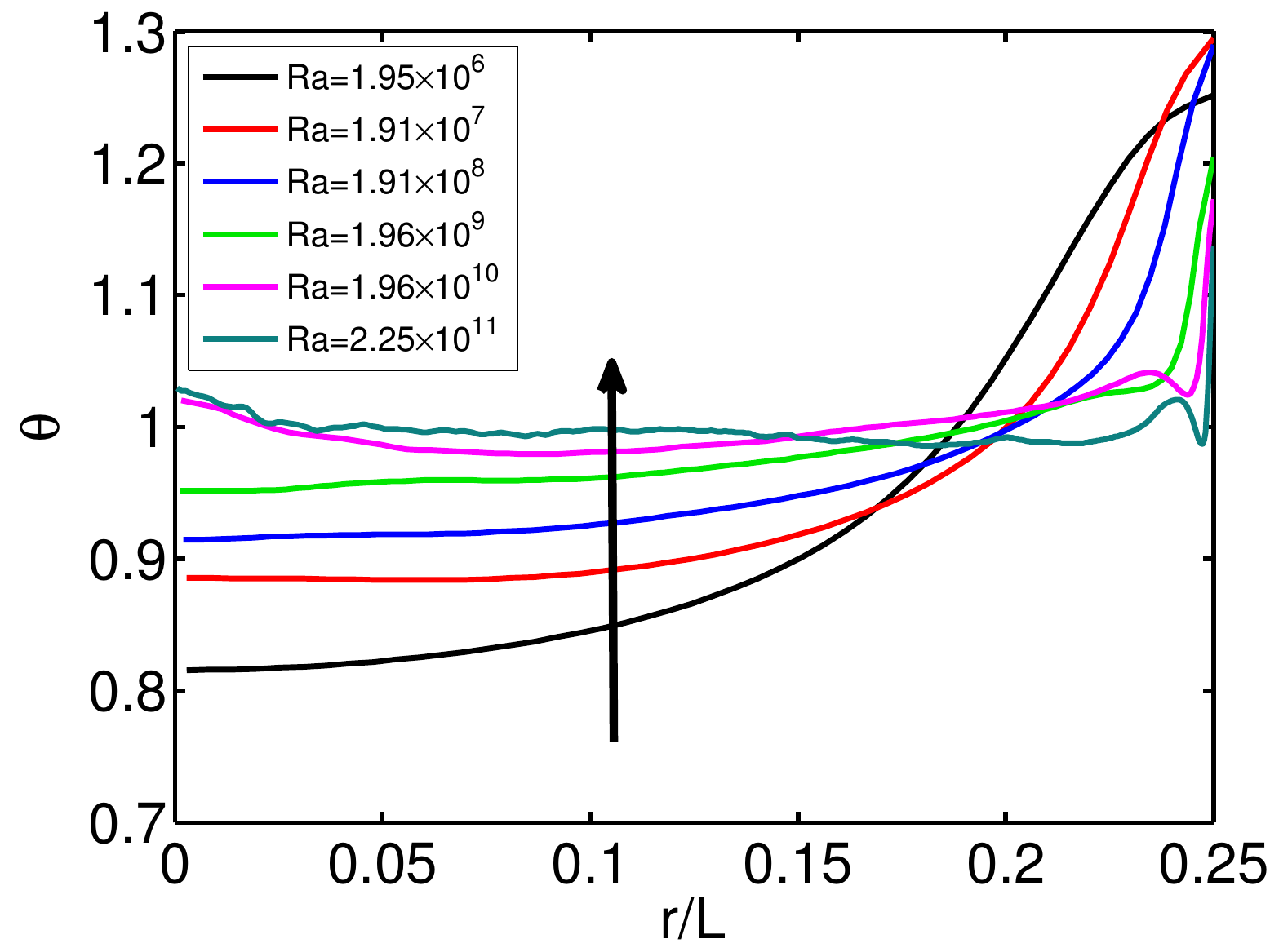}}
\caption{a) $Nu$ vs $Ra$ for $\Gamma=0.5$. Data as in figure \ref{fig:figure2}a. The numerical data for $Pr=0.7$ with constant heat flux at the bottom and constant temperature at the top plate are indicted in blue and the numerical data with constant temperature condition at both plates in black. b) Time-averaged temperature at the bottom plate for simulations with constant heat flux at the bottom plate and  constant temperature at the top plate. The arrow indicates the direction of increasing $Ra$.
}
\label{fig:figure4}
\end{figure}

It has also been argued that the different boundary conditions at the bottom plate, i.e.\
 that some experiments are closer to a constant temperature boundary condition, and some are closer to a constant heat flux boundary condition, might explain the differences in the heat transport that are observed in the high $Ra$ number regime. Figure \ref{fig:figure4}a compares the $Nu$ number in the simulations with 
 constant temperature and constant heat flux at the bottom plate. The figure shows that the difference between these both cases is small and even 
decreases with increasing $Ra$. For large $Ra$ no difference at all is seen within the (statistical) error bars,
which however increase due to the shorter averaging time (in terms of large eddy turnovers) at the
very large  $Ra \ge 2\times10^{11}$.

Figure \ref{fig:figure4}b shows the time-averaged temperature of the bottom plate in the simulations with 
constant heat flux at the bottom plate for different $Ra$. The radial dependence at the lower $Ra$ numbers can be understood from the flow structure in the sample:
Due to the large scale circulation the fluid velocities are largest in the middle of the sample. Thus in the middle more heat can be extracted from the plate than close to the sidewall where the fluid velocities are smaller. It is the lack of any wind in the corners of the sample that 
causes the relative high time-averaged plate temperature there. 
The figure also shows that this effect decreases with increasing $Ra$. The reason for this is that the turbulence becomes stronger  at higher $Ra$ and this leads to smaller flow structures. Therefore the region close to the sidewall with relative small fluid velocities decreases with increasing $Ra$ and this leads to a more uniform plate temperature at higher $Ra$. This effect explains that the simulations with constant temperature and constant heat flux condition at the bottom plate become more similar with increasing $Ra$.

The small differences between the simulations with constant temperature and 
 constant heat flux at the bottom plate shows that the differences between the experiments in the high $Ra$ number regime can not be explained by 
different plate conductivity properties. This finding is in agreement with the results of \cite{joh09}. In their periodic two-dimensional RB simulations the heat transfer for simulations with constant temperature and constant heat flux (both at the bottom and the top plate) 
becomes equal at $Ra \approx 5 \times 10^6$. For the three-dimensional simulations the heat transfer for both cases also becomes equal, but  at  higher $Ra$. This is due to the geometrical effect discussed before, see figure \ref{fig:figure4}b, that cannot occur in periodic two-dimensional simulations (\cite{joh09}).

\section{Conclusions}
In summary, we presented results from three-dimensional DNS simulations for RB convection in 
 cylindrical samples of aspect ratios $\Gamma=0.23$ and $\Gamma=0.5$ up to $Ra=2\times10^{12}$ and a broad range of $Pr$ numbers. The simulation at $Ra=2\times10^{12}$ with $Pr=0.7$ in an aspect ratio $\Gamma=0.5$ sample is in good agreement with the experimental results of \cite{nie00,nie01}, \cite{nie06}, \cite{fun09}, and \cite{ahl09c,ahl09d}, while there is a visible difference with the results of \cite{cha01}. In addition,
 we showed that the differences in the heat transfer observed between experiments for $Ra \gtrsim 2\times10^{11}$ can neither be explained by $Pr$ number effects, nor by the assumption of constant heat flux conditions at the bottom plate instead of constant temperature conditions. Furthermore, we demonstrated that the scaling of the kinetic and thermal BL thicknesses in this high $Ra$ number regime is well described by the Prandtl-Blasius theory.
 
Several questions remain: Which effect is responsible for the observed difference in $Nu$ vs $Ra$ scaling in the various experiments? Are there perhaps different turbulent states in the highly turbulent regime as has been suggested for RB flow by \cite{gro11}, but also for other turbulent flows in closed systems by \cite{cor10}? At what $Ra$ number do the BLs become turbulent? As in DNSs  both the velocity and temperature fields are known  in the whole domain (including in the boundary layers where the transition between the states is suggested to take place),  they will play a leading role in answering these questions.

\begin{acknowledgments}
\noindent
{\it Acknowledgement:} We thank J.\ Niemela, K.R.\ Sreenivasan, G.\ Ahlers, and P.\
Roche for providing the experimental data. The simulations were performed on Huygens (DEISA project), CASPUR, and HLRS. We gratefully acknowledge the support of Wim Rijks (SARA) and we thank the DEISA Consortium (www.deisa.eu), co-funded through the EU FP7 project RI-222919, for support within the DEISA Extreme Computing Initiative. RJAMS was financially supported by the  Foundation for Fundamental Research on Matter (FOM). We thank the Kavli institute, where part of the work has been done, for its hospitality. This research wasÊ supported in part by the National Science Foundation under Grant No.ÊPHY05-51164, via the Kavli Institute of Theoretical Physics.
\end{acknowledgments}


\end{document}